\documentclass[conference,anonymous]{IEEEtran}
\IEEEoverridecommandlockouts
\usepackage{cite}
\usepackage{amsmath,amssymb,amsfonts}
\usepackage{algorithmic}
\usepackage{bm}
\usepackage{enumitem}
\usepackage{graphicx}
\usepackage{multirow}
\usepackage{textcomp}
\usepackage{xcolor}
\usepackage{subcaption}
\usepackage{hyperref}
\def\BibTeX{{\rm B\kern-.05em{\sc i\kern-.025em b}\kern-.08em
    T\kern-.1667em\lower.7ex\hbox{E}\kern-.125emX}}
\begin{document}

\title{Improving Next-Application Prediction with Deep Personalized-Attention Neural Network
}

\author{\IEEEauthorblockN{Jun Zhu, Gautier Viaud, Céline Hudelot}
\IEEEauthorblockA{\textit{Mathématiques et Informatique pour la Complexité et les Systèmes} \\
\textit{CentraleSupélec, Université Paris-Saclay} \\
Gif-sur-Yvette, France \\
\{jun.zhu, gautier.viaud, celine.hudelot\}@centralesupelec.fr}
}

\maketitle

\begin{abstract}
Recently, due to the ubiquity and supremacy of E-recruitment platforms, job recommender systems have been largely studied. In this paper, we tackle the next job application problem, which has many practical applications. In particular, we propose to leverage next-item recommendation approaches to consider better the job seeker's career preference to discover the next relevant job postings (referred to jobs for short) they might apply for. Our proposed model, named Personalized-Attention Next-Application Prediction (PANAP), is composed of three modules. The first module learns job representations from textual content and metadata attributes in an unsupervised way. The second module learns job seeker representations. It includes a personalized-attention mechanism that can adapt the importance of each job in the learned career preference representation to the specific job seeker's profile. The attention mechanism also brings some interpretability to learned representations. Then, the third module models the \textit{Next-Application Prediction} task as a top-$K$ search process based on the similarity of representations. In addition, the geographic location is an essential factor that affects the preferences of job seekers in the recruitment domain. Therefore, we explore the influence of geographic location on the model performance from the perspective of negative sampling strategies. Experiments on the public CareerBuilder12 dataset show the interest in our approach.
\end{abstract}


\section{Introduction}
\label{sec:introduction}

Recently, the rapid development of E-recruitment has considerably influenced the human resource management field. In particular, due to the explosion of recruitment data (i.e., over 3 million jobs are posted on LinkedIn in the U.S. every month\footnote{\url{https://economicgraph.linkedin.com/resources/linkedin-workforce-report-february-2021}}), many works have been proposed to build effective job recommender systems~\cite{tripathi2016review}. Like many other domains, Deep Learning (DL) based recommendation models have been extensively studied in the recruitment domain. Nevertheless, while mostly studied, there are still important issues with current approaches, which we summarize below:
\begin{itemize}[leftmargin=18pt]
    \item \textbf{Issue 1}: The first issue is related to intrinsically dynamic and constantly evolving Job-Person interactions, which can be built from job application records or working histories. Identically, we assume that temporal relations between application records tend to point towards sequential/session-based recommendations\footnote{In this paper, we describe the recommendation task to explore the sequential data by a broader term ``sequential''. Thus ``session'' and ``sequential'' are interchangeable.}. To our knowledge, although various session-based methods have been proposed in others domains, session-based job recommendation is less studied. Moreover, another important aspect of the Job-Person interaction matrix is its sparsity, since most job seekers only apply for a few job positions or work as a specific occupation. 
    \item \textbf{Issue 2}: The second issue is related to the personalization of recommendations. Indeed, two job seekers can apply for the same job with different motivations. As a consequence, the important information in a job is specific to the job seeker. Thus, it is essential to weigh the jobs differently in the modeling tasks of job seeker careers. General-purpose approaches are often unable to capture such specific information. In addition, the personal context information of the job seeker is of prime importance in the recruitment domain. For example, the geographic location of a job seeker often has a substantial impact on the application decision and the recommendation result. It raises two opposite goals: (i) it is essential to consider this personal context information when modeling personal preference, (ii)  it also requires the recommender system to distinguish this information from more core content, i.e., job content, when making recommendations. 
    \item \textbf{Issue 3}: The last issue is due to the evolving nature of the recruitment domain. In fact, in addition, to constantly updated E-recruitment websites (i.e., a large number of job postings are added or removed daily, and hundreds of candidate profiles are created or updated), the domain is still evolving (i.e., new occupations, formations, and skills). All of this intensifies the cold-start problem. Therefore, it is impractical to build a static method that cannot be adjusted quickly as the data constantly changes. We thus need to avoid retraining a deep model for each change. Moreover, from the perspective of content analysis methods, due to the sensitivity and privacy of the data in the recruitment domain, it is not easy to create an unbiased and evolving labeled dataset~\cite{qin2018enhancing,zhu2018person}. Thus it prevents the use of supervised machine learning methods.
\end{itemize}
In this paper, we address the problem of job recommendation under the task of the next job application prediction. We propose a hybrid Personalized-Attention Next-Application Prediction model (PANAP) to answer the previous issues partially.  Specifically, for \textbf{Issue 1}, we model the next job application problem as a sequential recommendation problem, and then compare our proposed model with different session-based recommendation methods to analyze the dynamics in the recruitment domain. Our model is composed of three independent modules. The first module learns job representations from textual job content and available metadata in an unsupervised way to answer \textbf{Issue 3}. The second module contains a personalized-attention mechanism to learn the job seeker representation that can solve the problem of personalized recommendations mentioned in \textbf{Issue 2}. The third module is used to predict the next-application that a job seeker will apply for. In order to answer Issue 3, this module is inspired by the Deep Structured Semantic Model (DSSM)~\cite{huang2013learning} with a training loss based on representation similarities. Moreover, based on the nature of the recruitment domain, we propose a specific negative sampling strategy considering the geographic location factor. To summarize our contributions as follows:
\begin{itemize}[leftmargin=18pt]
    \item A new session-based model, named PANAP, is used for \textit{Next-Application Prediction} task. It integrates the personalized-attention mechanism to improve the recommendation accuracy.
    \item An extensive experimental study has been conducted in the CareerBuilder12 dataset, allowing us to investigate the effect of leveraging different metadata types and textual job contents, and the impact of different sampling strategies on the quality of recommendations.
\end{itemize}

\section{Related Work}
\label{sec:related works}
Job recommender systems based on traditional approaches (i.e., collaborative-, content-based and hybrid methods) have been well-studied~\cite{al2012survey}. However, these methods have some important limitations in the recruitment domain. For example, the sparsity of Job-Person interaction matrices limits the performance of collaborative-based methods. When the information contained in job contents is insufficient, or the feature-engineering on contents is difficult, content-based approaches may become inaccurate. Furthermore, these methods generally lack the ability to model personal information. Recently, DL has dramatically revolutionized recommendation systems. As in other domains, DL-based methods have been applied in the recruitment domain. These works can be categorized into three classes according to the data they used and the learning paradigm. (i) \textbf{Text-based matching models}: When labeled Person-Job matching records are available, the recommendation problem can be seen as a supervised text matching task to match jobs with candidates automatically. Many approaches~\cite{qin2018enhancing,bian2019domain,zhu2018person} with different architectures have been proposed in the literature. However, due to the protection of trade secrets and personal privacy, collecting such labeled records is difficult. (ii) \textbf{Unsupervised representation learning models}: When matching records are not available, the recommendation task can be seen as the learning of representations of job and person in the same embedding space. The recommendation is then modeled as a top-$K$ search based on these learned embeddings using similarity-based algorithms. For example, ~\cite{dave2018combined} uses graphs to learn job title and skills representations. ~\cite{zhang2019job2vec} proposes a collective multi-view method to learn job title representations. In these approaches, the different level/view of information is of prime importance. (iii) \textbf{Approaches based on career paths}:
The methods mentioned above primarily focus on characterizing the person or the job information. Transitions in career paths are less considered. To address this problem, some works are proposed, e.g., NEMO~\cite{li2017nemo} explores action dependencies in career paths through Long Short-Term Memory (LSTM)~\cite{hochreiter1997long} to predict the next career move. In~\cite{meng2019hierarchical}, they use a hierarchical LSTM to predict the next potential employer of a person and how long he/she will stay in the new position. However, predicting the next job application based on application records received relatively little consideration except~\cite{lacic2020using}, which recommends the next job posting in a $K$-nearest neighbor manner. It is one of our baselines.

The next-application prediction problem can be considered as a sequential recommendation problem, which has been extensively studied recently in other domains, such as news or product recommendation~\cite{de2018news,li2017neural,yu2016dynamic,fang2020deep}. Recurrent Neural Networks (RNNs) have been widely studied for the sequential recommendation since they have demonstrated their effectiveness in processing sequential data, one representative work is GRU4Rec~\cite{hidasi2015session,hidasi2016parallel,hidasi2018recurrent}. The attention mechanism~\cite{bahdanau2014neural} has shown promising potential in improvements of accuracy and interpretability, like the vanilla attention-based~\cite{li2017neural,liu2018stamp} and self-attention based one~\cite{sun2019bert4rec,kang2018self}.

Different from the previous state-of-the-art in the recruitment domain, in this paper, we tackle the problem of next-application prediction by leveraging session-based recommendation models. We further study the effectiveness of session-based models in the recruitment domain. Moreover, we integrate the personalized-attention mechanism to model the different informativeness of job postings for different job seekers, and we take into account the ``location'' factor when making recommendations.

\section{Problem Formulation}
\label{sec:problem}
Let $\mathcal{U}$ be a set of job seekers and $\mathcal{J}$ be a set of job positions (referred to jobs for short). A job $j\in\mathcal{J}$ can be represented as $j=(\text{ID}_{j}, \{w_{1}^{j},\dots, w_{m}^{j}\}, \text{meta}_{j})$, where $\text{ID}_{j}$ is the unique identifier of job $j$. $\{w_{1}^{j},\dots, w_{m}^{j}\}$ is a sequence of $m$ words, representing the textual content of job $j$. It can be any text information about the job. $\text{meta}_{j}$ are associated metadata attributes (i.e., job city). Similarly, each job seeker $u\in\mathcal{U}$ can be summarized as a tuple $u=(\text{ID}_{u}, \mathcal{H}_{u}, \text{meta}_{u})$ with an identifier $\text{ID}_{u}$, a professional profile $\mathcal{H}_{u}$, and metadata attributes $\text{meta}_{u}$, like the education background or the geographical information. Specifically, the professional profile $\mathcal{H}_{u}$ can be represented either as his/her working experience or his/her job application record. In this work, we target at the job application record, and we denote $\mathcal{H}_{u}$ as a job sequence (session) ordered by time $\mathcal{H}_{u} = \{j_{1}, \dots, j_{n}\}$, where each $j_{i}$ is a specific job in $\mathcal{J}$. With the above notations, the \textit{Next-Application Prediction} problem can be formally defined as follows:

\textbf{Given} a set of jobs $\mathcal{J}$ and a set of job seeker $\mathcal{U}$, for each specific job seeker $u\in\mathcal{U}$ who has a job application sequence $\mathcal{H}_{u} = \{j_{1}, \dots, j_{t}\}$ at one specific time step $t$.

\textbf{Predict} the next most likely job $j_{t+1}$ that the job seeker $u$ might apply for, which means maximizing the likelihood $P(j_{t+1} = j^{+}|\mathcal{H}_{u},\mathcal{J})$ of the actual next-applied job $j^{+}$ given the application sequence $\mathcal{H}_{u}$.

\section{The Proposed Model}
\label{sub:model}
An overview of our PANAP model is given in Figure 1. We detail our model by describing these three different modules.
 \begin{figure*}[h]
 \centerline{\includegraphics[width=1.\textwidth]{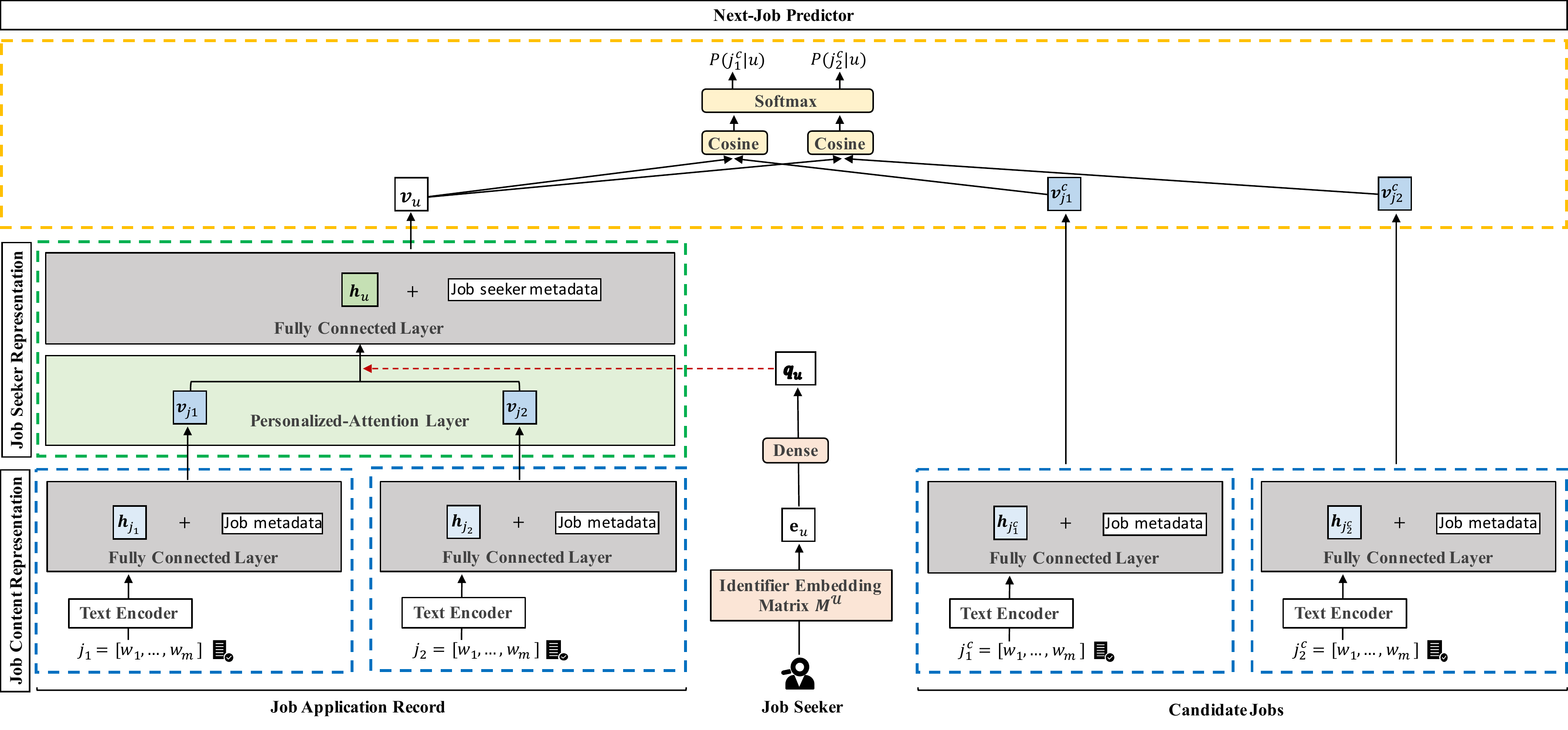}}
  \caption{The proposed PANAP framework consists of three parts: (i) \textit{Job Content Representation} (blue dashed box) is used for job content representation learning. (ii) \textit{Job Seeker Representation} (green dashed box) uses the personalized-attention mechanism to characterize the career preference of job seekers based on the Job Application Record. (iii) \textit{Next-Application Predictor} (yellow dashed box) utilizes a ranking loss based on the representation similarity of job and job seeker to train the model.}
  \label{fig:structure}
\end{figure*}

\subsection{Job Content Representation}
JCR (Job Content Representation), is responsible for learning a representation of each job $j\in\mathcal{J}$, where $j=(\text{ID}_{j}, \{w_{1}^{j},\dots, w_{m}^{j}\}, \text{meta}_{j})$. Specifically, for job $j$, a textual representation $\bm{h}_{j}\in \mathbb{R}^{d}$ is learned from its textual content $\{w_{1}^{j},\dots, w_{m_{j}}^{j}\}$ with a text encoder: 
\setlength{\abovedisplayskip}{1.3pt}
\setlength{\belowdisplayskip}{1.3pt}
\begin{equation}
\bm{h}_{j} = \text{text\_encoder}(\{w_{1}^{j},\dots, w_{m}^{j}\}),
\end{equation}
where the text encoder can be some unsupervised textual embedding approaches (e.g., Word2Vec~\cite{mikolov2013efficient} and Doc2Vec~\cite{le2014distributed}) or pre-trained models (e.g., BERT~\cite{devlin2018bert}). We believe that textual representations generated by different text encoders will affect the accuracy of the recommendations. It will a research line in our future work. The metadata attributes $\text{meta}_{j}$  (e.g., city and state) are respectively embedded into vectors (e.g., $\bm{v}_{j}^{\text{city}}$ and $\bm{v}_{j}^{\text{state}}$) through different trainable embedding matrices (e.g., $\bm{M}^{\text{city}}$ and $\bm{M}^{\text{state}}$), which will be jointly learned during the training process. The vector $\bm{v}_{j}^{\text{meta}}$ is then obtained by concatenation of all vectors, $\bm{v}_{j}^{\text{meta}} = [\bm{v}_{j}^{\text{city}}\oplus \bm{v}_{j}^{\text{country}}\oplus\dots]$, where the symbol $\oplus$ represents the concatenation operator. Then the textual job representation $\bm{h}_{j}$ and job metadata vector $\bm{v}_{j}^{\text{meta}}$ are combined by using a sequence of Fully Connected (FC) layers to produce the \textit{Job Content Vector} $\bm{v}_{j}\in\mathbb{R}^{d^{\mathcal{J}}}$:
\begin{equation}
    \bm{v}_{j}=\text{FCs}\big([\bm{h}_{j}\oplus\bm{v}_{j}^{\text{meta}}]\big).
\end{equation}
It is worth mentioning that the textual job representation $\bm{h}_{j}$ is trained separately in an unsupervised way or generated by a pre-trained language model, so that a new job can be added easily, thereby alleviating the job cold-start problem.

\subsection{Job Seeker Representation}
JSR (Job Seeker Representation), is responsible for learning representations of job seekers. It requires three inputs: the identifier $\text{ID}_{u}$ of job seeker $u$, his/her historical applied job sequence $\mathcal{H}_{u} = \{j_{1}, \dots, j_{n}\}$\footnote{We omit the superscript of $u$ without loss of clarity.} and associated metadata attributes $\text{meta}_{u}$. In addition, $\mathcal{H}_{u}$ can be transformed as a sequence of \textit{Job Content Vector}s $\{\bm{v}_{j_{1}}, \dots, \bm{v}_{j_{n}}\}$, where each $\bm{v}_{j_{i}}\in\mathbb{R}^{d^{\mathcal{J}}}$ is obtained from JCR module. Since the same job may have different informativeness for different job seekers, it contributes differently to characterize career profiles of job seekers. Then, we learn the job seeker representation with a personalized-attention mechanism~\cite{wu2019npa}. We first embed the $\text{ID}_{u}$ of job seeker $u$ into a vector $\bm{e}_{u}$ using an identifier embedding matrix $\mathbf{M}^{\mathcal{U}}\in\mathbb{R}^{\vert\mathcal{U}\vert\times d^{s}}$, where $d^{s}$ denotes the dimension of identifier embedding. Then $\bm{e}_{u}$ is passed to a dense layer parameterized with $\bm{W}^{q}\in \mathbb{R}^{d^{\mathcal{U}}\times d^{q}}$ and $\bm{b}^{q}\in \mathbb{R}^{d^{q}}$ to form the preference query vector $\bm{q}_{u} = \text{ReLU}(\bm{W}^{q}\times\bm{e}_{u} + \bm{b}^{q})$, where $d^{q}$ is the query dimension. 
The importance score of $i$-th job for job seeker $u$ is calculated as follows:
\begin{equation}
     \alpha_{i} =\exp\big(\bm{v}_{j_{i}}^{T}\bm{p}_{u}\big)/ \textstyle\sum_{i^{\prime}=1}^{n}\exp{\big(\bm{v}_{j_{i^{\prime}}}^{T}\bm{p}_{u}}\big),
\end{equation}
where $\bm{p}_{u} =\text{tanh}\big(\bm{W}^{a}\times\bm{q}_{u}+\bm{b}^{a}\big)$, with projection parameters $\bm{W}^{a}\in \mathbb{R}^{d^{q}\times d^{\mathcal{J}}}$ and $\bm{b}^{a}\in\mathbb{R}^{d^{\mathcal{J}}}$. Note that the attention scores might differ for the same job, as the query vector depends on the ID of the job seeker. This allows forming a weighted career preference representation $\bm{h}_{u}$ for job seeker $u$:
\begin{equation}
    \bm{h}_{u} = \textstyle\sum_{i=1}^{n}\alpha_{i}\bm{v}_{j_{i}}.
\end{equation}
Similar to JCR module, the metadata vector of job seeker $u$ is represented as $\bm{v}_{u}^{\text{meta}} = [\bm{v}_{u}^{\text{city}}\oplus \bm{v}_{u}^{\text{education}}\oplus\dots]$ by concatenation. $\bm{v}_{u}^{\text{meta}}$ and $\bm{h}_{u}$ are combined by using a sequence of FC layers to produce the final job seeker representation:
\begin{equation}
    \bm{v}_{u}=\text{FCs}\big([\bm{h}_{u}\oplus\bm{v}_{u}^{\text{meta}}]\big),
\end{equation}
where $\bm{v}_{u}$ has a dimension of $d^{\mathcal{U}}$, which is equal to 
the dimension of \textit{Job Content Vector} $\bm{v}_{j}$ (e.g., $d^{\mathcal{U}}=d^{\mathcal{J}}$).

\subsection{Next-Application Predictor}
\label{sub:next-application predictor}
In classical deep recommendation models, the output is usually a probability vector whose dimension is the number of available items. In our scenario of job recommendation, due to the evolving nature of the recruitment domain (Issue 3), such models are not tractable in practice. Thus this dynamic scenario needs an approach that does not need to retrain the whole network at each change. As a consequence, inspired by the highly dynamic news recommendation scenario, we use the ranking loss~\cite{de2018news} to train the predictor. The principle of this loss is to train the predictor to maximize the similarity between user preferences and positive samples, while minimizing similarities with negative samples. The idea of the ranking loss comes from DSSM~\cite{huang2013learning}, which is an effective document ranking model and has been leveraged for the recommendation. In our case, it can immediately recommend a newly published job as soon as its representation is learned. More precisely, given the application history $\mathcal{H}_{u} = \{j_{1}, \dots, j_{n}\}$ of job seeker $u$, we formulate the above representation generating processes as $\bm{v}_{u} = JSR(\text{ID}_{u}, \mathcal{H}_{u},\text{meta}_{u};\Theta_{1})$, where each $j_{i}\in\mathcal{H}_{u}$ can be embedded into a \textit{Job Content Vector} $\bm{v}_{j_{i}}$ through $JCR(j_{i}; \Theta_{2})$, $\Theta_{1}$ and $\Theta_{2}$ are model parameters. $\bm{v}_{u}$ and $\bm{v}_{j_{i}}$ are vectors of the same dimension, and we thus can define the relevance score between job seeker $u$ and job $j$ by the cosine similarity of their representations:
\begin{equation}
    sim(u,j_{i}) = cos(\bm{v}_{u},\bm{v}_{j_{i}}) = \frac{\bm{v}_{u}\cdot\bm{v}_{j_{i}}}{\|\bm{v}_{u}\|\|\bm{v}_{j_{i}}\|}
\end{equation}

The posterior probability of $j$ being the next-job for $u$ is formulated as follows:
\begin{equation}
    P(j|u) = \frac{\exp\big(sim(u,j)\big)}{\sum_{j^{\prime}\in{j^{+}\cup \mathcal{J}^{-}}}\exp\big(sim(u,j^{\prime})\big)},
\end{equation}
where $j^{+}$ is the actual next-application, and $\mathcal{J}^{-}$ represents the negative sample set of jobs which are not applied by the job seeker during his active session. The recommender should learn to maximize the similarity between the content vector $\bm{v}_{j^{+}}$ of $j^{+}$ and the job seeker preference vector $\bm{v}_{u}$, while minimizing similarities with job vector $\bm{v}_{j^{-}}$ in the set $\mathcal{J}^{-}$.
\begin{equation}
    \mathcal{L}_{sim} = -\log \textstyle\prod_{(u, j^{+})} P(j^{+}|\text{ID}_{u}, \mathcal{H}_{u},\text{meta}_{u};\Theta_{1},\Theta_{2}).
\end{equation}

\subsection{Negative Sampling Strategies}
\label{sub:Negative Sampling Strategies}
Since our method is trained and evaluated with negative samples as described in Section~\ref{sub:next-application predictor}, the negative sample set $\mathcal{J}^{-}$ significantly influences the model performance. A well-used sampling strategy is the mini-batch based sampling proposed in~\cite{hidasi2015session}, which treats the items from the other training/evaluation sessions in the same mini-batch as negative samples. ~\cite{hidasi2018recurrent} extends the mini-batch based strategy by adding additional samples (based on unity or based on popularity). For some applications, such as news, music, and video, popularity is also an essential factor that influences user choice besides personal preference. Therefore, sampling based on popularity is a good sampling strategy. However, in the recruitment field, unlike these applications, the choices of job seekers are more influenced by their personal contexts, such as the geographic location factor. When job seekers make a choice, they usually first need to consider the job location. They are more likely to apply for jobs in their cities or other cities not far from their current locations (i.e., cities in the same state). We will prove this observation in Section~\ref{sec:datasets}. To solve this problem, our method cooperates with the ``location'' metadata attributes (e.g., city, state and country) of job and job seeker to model their representations, respectively. As a consequence, these representations contain the ``location'' information. In addition, considering the ``location'' when generating negative samples can enable our model to learn useful information for more meaningful recommendations. More experimental details are provided in Section~\ref{sub:Experimental Settings}.

\section{Experiments}
\label{sec:experiments}

\subsection{Datasets}
\label{sec:datasets}
We employ \textit{CareerBuilder12}\footnote{\url{https://www.kaggle.com/c/job-recommendation}} to evaluate our proposed method. Its statistics are given in Table~\ref{tab:datasets}. In this dataset, job seeker has five metadata: \textit{City}, \textit{State}, \textit{Country}, \textit{Degree}, and \textit{Major}. Job metadata are \textit{City}, \textit{State} and \textit{Country}. The textual job content includes a \textit{Job Title}, a \textit{Job Description} and some \textit{Job Requirements}. From this initial dataset, we created two datasets: (i) \textit{CB12\_s} like in~\cite{lacic2020using}, in which sessions are created via a time-based split of 30 minutes inactivity threshold, and we discarded sessions with less than two applications for next-job prediction purpose. (ii) \textit{CB12\_l} uses all application records during 13 weeks to model the career profile of each job seeker. Thus, \textit{CB12\_l} has longer sequences that enable us to evaluate the effectiveness of our proposed method. We further split the last 14 days for testing and the remaining sessions for training. We filter job applications in the test set that do not belong to the training set as this enables a better comparison with the approaches, which can only recommend items that have been used to train the model. 
\begin{table*}[htbp]
  \caption{Statistics of datasets, \textit{$\#$S} represents the session number, and \textit{$\#$A} is the applications number. \textit{Avg\_SLen} is the average application number in sessions. $\#$\textit{Metadata} contains the cardinality of each metadata attribute.}
  \begin{center}
  \begin{tabular}{|c|cccccc|}
  \hline
  \multirow{2}{*}{\textbf{Dataset}}  & \multirow{2}{*}{$\vert\mathcal{U}\vert$}
    &  \multirow{2}{*}{$\vert\mathcal{J}\vert$} &  \multirow{2}{*}{$\#$S}  &  \multirow{2}{*}{$\#$A} & \multirow{2}{*}{Avg\_SLen} &  \multirow{2}{*}{\textbf{$\#$Metadata}}
    \\[0.5em]
    \cline{7-7}
    & & & & & &
    City/State/Country/Degree/Major \\
    \hline
   CB12\_s & 111,785 & 207,972 &  165,027 &  638,469 & 3.87 & 8,226/122/33/7/21,224  \\
  CB12\_l & 137,642 & 239,581 & 137,642 &  772,305 & 5.61 & 8,856/130/37/7/25,201 \\
  \hline
\end{tabular}
\label{tab:datasets}
\end{center}
\end{table*}

\begin{figure}
    \centering
    \begin{subfigure}[b]{0.45\textwidth}
     \centerline{\includegraphics[width=0.8\textwidth]{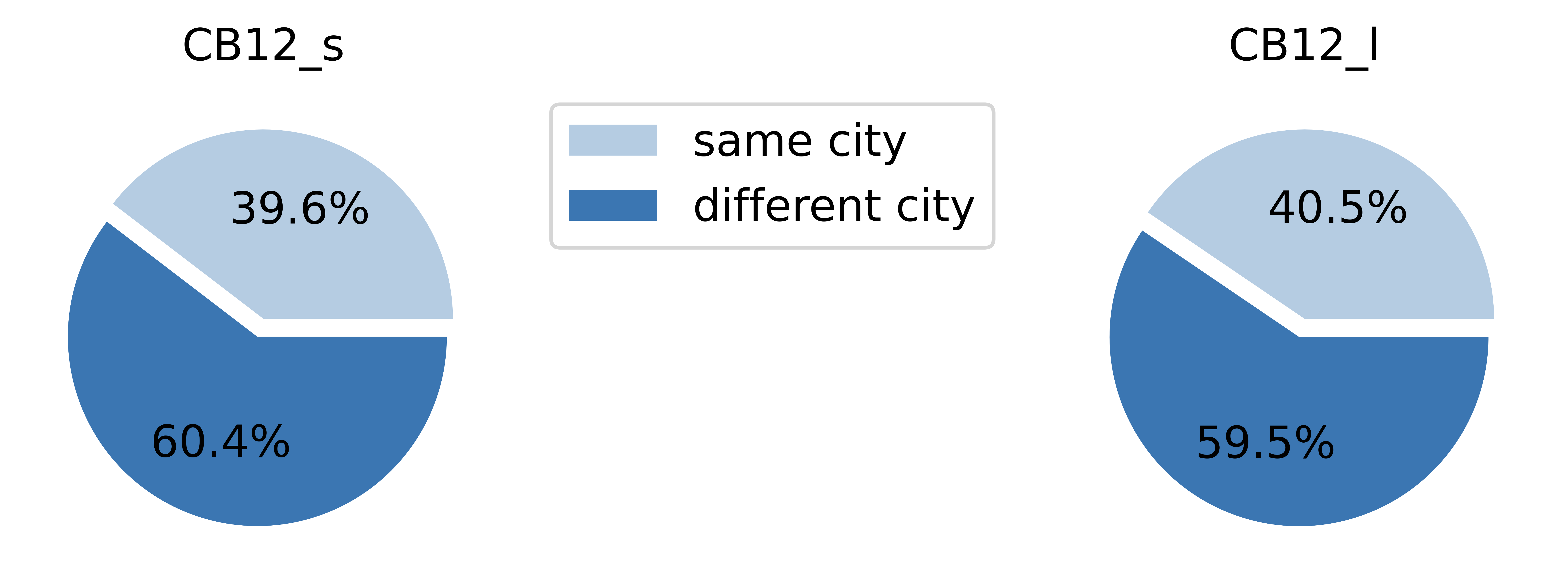}}
    \vspace{0.001cm}
    \caption{City}
    \label{fig:city}
    \end{subfigure}
    \hspace{0.1cm}
    \begin{subfigure}[b]{0.45\textwidth}
     \centerline{\includegraphics[width=0.8\textwidth]{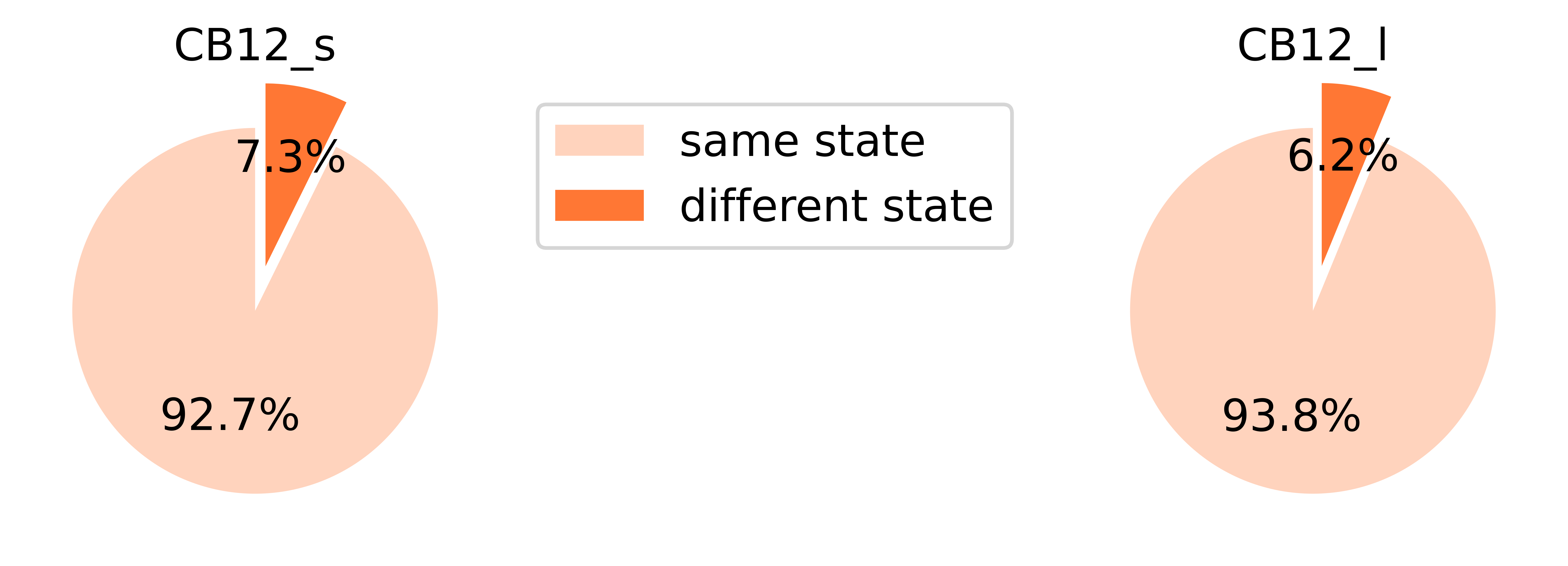}}
    \vspace{-1.cm}
    \caption{State}
    \label{fig:state}
  \end{subfigure}
  \caption{The relationship between locations of the job seeker and the applied job in \textit{CareerBuilder12} datasets.}
  \label{fig:choice}
\end{figure}
\setlength{\textfloatsep}{5pt}

As we described in Section~\ref{sub:Negative Sampling Strategies}, the ``location'' is an essential factor that needs to be considered in the job recommendation. In order to further prove this observation, in Figure \ref{fig:choice}, we illustrate the relationship between locations of job seekers and applied jobs in \textit{CareerBuilder12} datasets. As shown in Figure \ref{fig:state}, most job seekers ($92.7\%$ in CB12\_s and $93.8\%$ in CB12\_l) have applied for jobs in their states. Among these people, $81.3\%$ and $79.7\%$ of job seekers only consider jobs in their own states. Figure \ref{fig:city} shows that only $39.6\%$ and $40.5\%$ of job seekers apply for jobs in their own cities, because job opportunities in their cities are usually limited. People are also applying for jobs in other cities in the same state, as explained above. Therefore, the ``location'' (the job site or the current location of the job seeker) is an essential factor to be considered in the job recommendation.

\subsection{Experimental Settings}
\label{sub:Experimental Settings}
In this experiment, job representations $\bm{h}_{j}$ were obtained by Doc2Vec with dimension $d=300$ via the distributed memory. The dimension $d^{s}$ of identifier embedding $\bm{e}_{u}$ and the query dimension $d^{q}$ were set to 100. The dimensions $d^{\mathcal{U}}$ and $d^{\mathcal{J}}$ were also set to 300. We applied the dropout technique with a rate of 0.2 to each layer and L2 regularization with rate 1e-4 to parameter weights. The PANAP was trained and evaluated with a 256 mini-batch size, and we used 15 negative samples for training and 50 for evaluation. The Adam~\cite{kingma2014adam} optimizer with a learning rate of 5e-4 was used. Metadata attributes with low cardinality ($<=10$) were one-hot encoded, and high cardinality attributes were represented as trainable embeddings. We used two FC layers, with Leaky ReLU~\cite{maas2013rectifier} and tanh activation functions to combine metadata vectors. 

We first explore a sampling strategy inspired by~\cite{gabriel2019contextual}, and then propose an improved sampling strategy for the job recommendation scenario. This improved strategy considers the current geographic location of the job seeker and the job site when sampling negative jobs. The two strategies are:
\begin{itemize}[leftmargin=18pt]
    \item \textbf{Strategy 1 (S1)}-\textit{mini-batch + additional samples}: It is proposed in~\cite{gabriel2019contextual}, which adds additional samples with a uniform sampling strategy from a global buffer of the $N$ most recently applied jobs, when there are not enough negative samples within the mini-batch. Jobs are uniformly sampled from the ``candidate set'' (i.e., jobs within the mini-batch and additional jobs);
    \item \textbf{Strategy 2 (S2)}-\textit{mini-batch + additional samples + location-biased}: Different from  Strategy 1, in Strategy 2, we first select the jobs in the same state as the job seeker from the “candidate set” as negative samples. When there are not enough negative samples in the current state, we sample jobs in other states from the ``candidate set''.
\end{itemize}

\begin{figure}
    \centering
    \begin{subfigure}[b]{0.21\textwidth}
     \centerline{\includegraphics[width=1.1\textwidth]{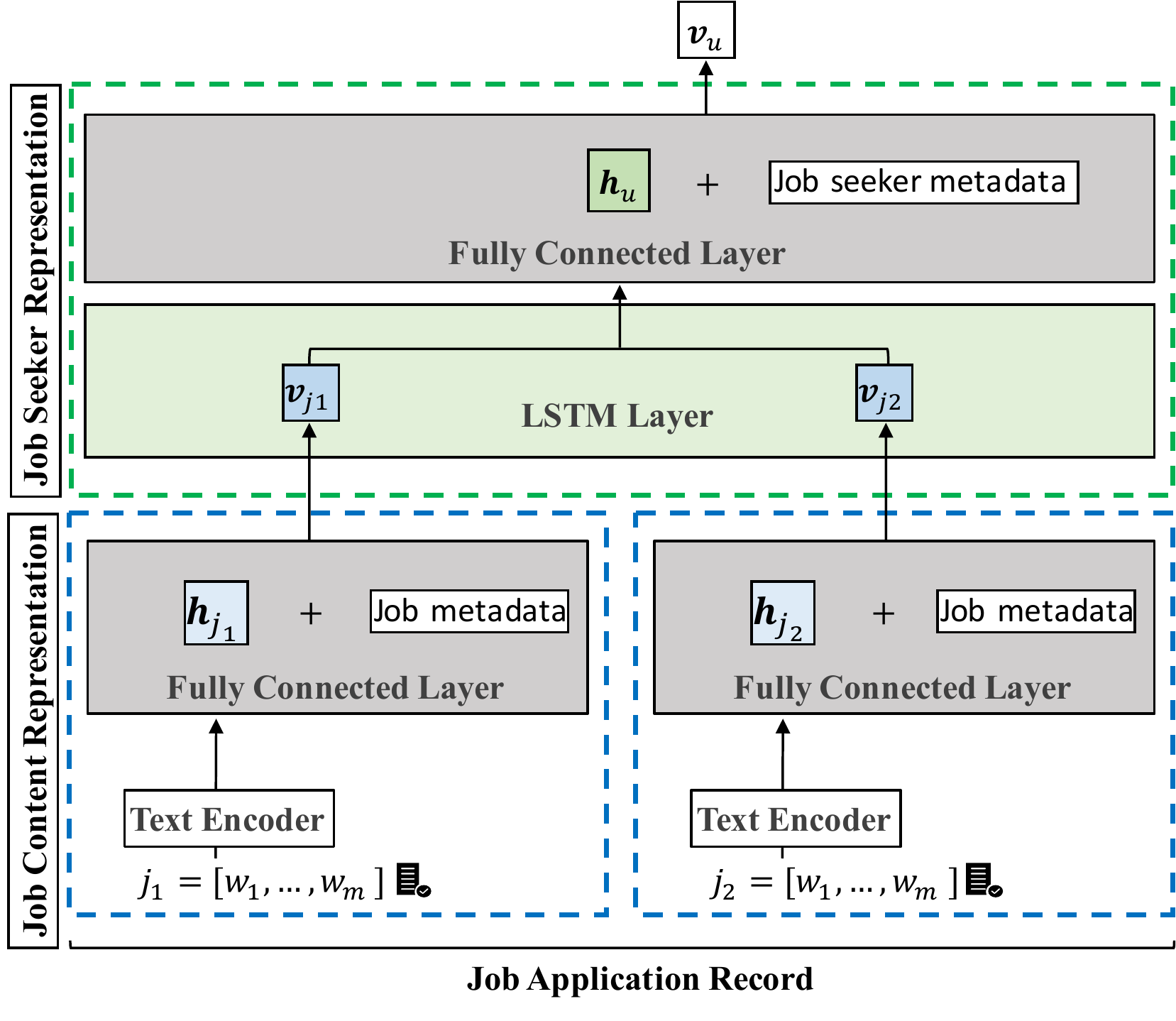}}
    \caption{LSTM}
    \label{fig:LSTM}
    \end{subfigure}
    \hspace{1.5em}%
    \begin{subfigure}[b]{0.21\textwidth}
    \centerline{\includegraphics[width=1.1\textwidth]{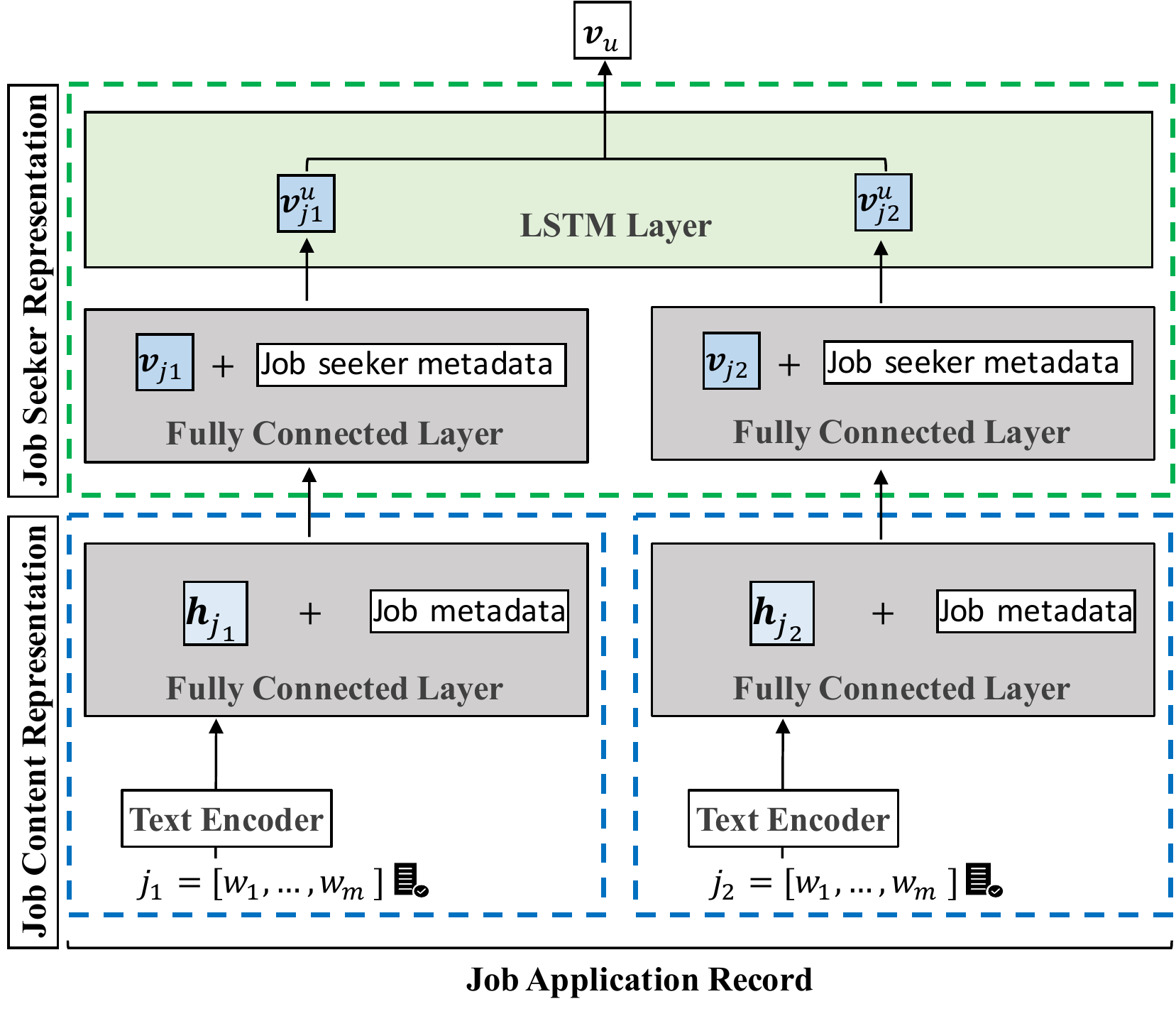}}
    \caption{PLSTM}
    \label{fig:PLSTM}
  \end{subfigure}
  \caption{Two variants of PANAP.}
\end{figure}

The baselines used are listed as follows:
\begin{itemize}[noitemsep,topsep=2mm,leftmargin=18pt]
     \item \textit{POP}: recommends the most applied job;
     \item \textit{Association Rule (AR)}: is a simplified version of association rule~\cite{agrawal1993mining} with a maximum rule size of two;
     \item \textit{Content Similarity (CS)}: recommends similar jobs based on the cosine similarity between representations of each applied job and the $N$ most recently applied jobs in the global buffer;
     \item \textit{Item-kNN (IkNN)}: recommends jobs that are similar to the last applied job during the current session as in~\cite{hidasi2015session};
     \item \textit{Session-kNN (SkNN)}~\cite{jannach2017recurrent}: compares the entire current session with the past sessions in the training dataset, rather than considering only the last job;
     \item \textit{Vector Multiplication SkNN (V-SkNN)}~\cite{ludewig2018evaluation}: is a variant of \textit{SkNN} that emphasizes jobs more recently interacted within the current session, when computing the similarities with past sessions (a linear decay function is used);
     \item \textit{VAE\_{Comb}}: is the best model proposed in~\cite{lacic2020using};
     \item \textit{GRU4Rec}: is the most recent version of \textit{GRU4Rec}~\cite{hidasi2018recurrent};
     \item \textit{Bert4Rec}~\cite{sun2019bert4rec}: introduces the bi-directional self-attention model to model user behavior sequences;
     \item Two variants of PANAP: (i) \textit{LSTM} replaces the personalized-attention layer with a LSTM layer, Figure \ref{fig:LSTM}, and (ii) \textit{PLSTM} adds job seeker metadata to each job representation to generate \textit{Personalized job embedding}~\cite{de2018news} before the LSTM layer, Figure \ref{fig:PLSTM}.
 \end{itemize}
The evaluation metrics used in this work are Hit Rate (HR@5)~\cite{ludewig2018evaluation}, Mean Reciprocal Rank (MRR@5)~\cite{ludewig2018evaluation} and Normalized Discounted Cumulative Gain (NDCG@5)~\cite{fang2019deep}.

\section{Results}
\label{sec:results}
In this section, we present our experimental results (HR@5/MRR@5/NDCG@5). For all tables shown in this section, the score in bold is the best in each metric, and the score in the underline is the second best.

\subsection{Next-Application Prediction}
\begin{table}[htbp]
\caption{\textit{Next-Application Prediction} performance.}
\begin{center}
\begin{tabular}{|c|cc|}
\hline
\small Method & CB12\_s & CB12\_l \\
\hline
\small POP & 0.089/0.042/0.054 & 0.093/0.046/0.058\\
\small AR & 0.273/0.200/0.218 & 0.202/0.146/0.160\\
\small CS & 0.226/0.128/0.152 & 0.214/0.118/0.142\\
\small IkNN & 0.274/0.202/0.220 & 0.202/0.147/0.161\\
\small SkNN & 0.349/0.247/0.272 & 0.276/0.196/0.216\\
\small V-SkNN & 0.349/0.248/0.273 & 0.276/0.196/0.217\\
\small VAE\_{Comb} & 0.345/0.239/0.256 & 0.282/0.203/0.224\\
\small GRU4Rec & 0.367/0.208/0.230  & 0.403/0.309/0.352 \\
\small BERT4Rec & 0.373/0.232/0.245 & 0.438/0.311/0.369\\
\hline
\small PANAP (S2) &  \textbf{0.691/0.492/0.541}  & \textbf{0.742/0.543/0.593}\\
\small LSTM & 0.540/0.339/0.389 & 0.569/0.368/0.418\\
\small PLSTM & 0.480/0.292/0.338 & 0.516/0.323/0.371 \\
\hline
\end{tabular}
\label{tab:prediction}
\end{center}
\end{table}
According to Table \ref{tab:prediction}, we have the following observations: (i) Among all models, \textit{POP} and \textit{CS} give the lowerest scores, as they neither model the personalized preference nor consider the sequential information. Although \textit{POP} is often a strong baseline in certain domains, i.e., news and movies, career preferences are less affected by popularity factors in the recruitment domain. (ii) Overall, the NN-based methods consistently outperform traditional methods, demonstrating that NNs are good at modeling sequential information, and the self-attention mechanism can improve accuracy. (iii) Our proposed method \textit{PANAP (S2)} with sampling strategy S2 and its variants \textit{LSTM} and \textit{PLSTM} perform best among all baselines, which indicates the job content and metadata can effectively improve the recommendation performance, similar observations can be found in Section \ref{sub:features}. (iv) \textit{PANAP (S2)} outperforms \textit{LSTM} and \textit{PLSTM}, which use LSTM to model the sequential information. One possible reason is that in the recruitment domain, career preferences are less dynamic than other domains, and application sequences are relative short (3.87 and 5.61 on average for both sets), thus the advantage of RNN can not be well demonstrated. This result also demonstrates the advantage of personalized-attention as different jobs might have different importance for career preference modeling, and selecting the more critical jobs is useful for achieving better recommendation performance. Moreover, \textit{LSTM} is better than \textit{PLSTM}, one possible reason is that \textit{PLSTM} merges the job seeker metadata into each job in the session, which will weaken the information carried by the job itself.


\subsection{Effectiveness of Personalized Attention}
In this part, we analyze the effectiveness of the personalized-attention mechanism using CB12\_s dataset. As shown in Table \ref{tab:attention}, the models with attention mechanism consistently outperform the model without attention, and our model with the personalized-attention outperforms its variant with vanilla attention (similar to the global attention used in~\cite{meng2019hierarchical}). Such a result is probably since the vanilla attention uses a fixed query vector and cannot adjust to different personal preferences.
\begin{table}[ht]
  \caption{Different attention mechanisms.}
  \begin{center}
  \begin{tabular}{|c|c|}
    \hline
    Attention\_mechanism & HR@5/MRR@5/NDCG@5\\
     \hline
    Personalized-attention &  \textbf{0.691/0.492/0.541} \\
    Vanilla attention & 0.685/0.480/0.531\\
    No attention (avg) & 0.565/0.367/0.416\\
    LSTM & 0.540/0.339/0.389 \\
   \hline
\end{tabular}
\label{tab:attention}
\end{center}
\end{table}

\subsection{Effectiveness of Different Features}
\label{sub:features}

As shown in Table~\ref{tab:meta}, (i) It is obvious that methods with the additional information (i.e., content representation or metadata) generally give better results than what is achievable from the job identifier $\textit{ID}_{j}$ only (\textit{Only\_JobID}) on CB12\_s dataset. (ii) The metadata has a fewer influence on \textit{PANAP (S2)} than \textit{LSTM} (29.1\% average reduction between \textit{Meta+Content+JobID} and \textit{No\_Meta} on three metrics compared to 41.0\%). One possible reason is that \textit{PANAP (S2)} utilizes a personalized-attention mechanism to model career preferences, which already contain personal information. (iii) Since the ``location'' factor affects the choice of job seekers, and our sampling strategy is location-biased. The job metadata, e.g., \textit{City}, \textit{State} and \textit{Country} could give more relevant information about ``location'', so the scores of \textit{No\_JobMeta} are lower than that of \textit{No\_SeekerMeta}. (iv) \textit{PANAP (S2)} consistently outperforms \textit{LSTM}, even only the job identifier $\textit{ID}_{j}$ is used, which indicates that \textit{PANAP (S2)} does capture personal preference. This observation also proves the advantage of the personalized-attention mechanism in cases where no additional information is available.

\begin{table}[htbp]
  \caption{Different feature combinations. \textit{No\_Meta} means that neither the job seeker metadata nor the job metadata is considered. }
  \begin{center}
  \begin{tabular}{|c|cc|}
    \hline
    Feature & PANAP (S2) & LSTM \\
    \hline
    Meta+Content+JobID & \textbf{0.691/0.492/0.541} & 0.540/0.339/0.389 \\
    No\_Meta & 0.516/0.334/0.379 & 0.341/0.189/0.226\\
    No\_Content & 0.530/0.341/0.386 & 0.449/0.286/0.326 \\
    Only\_JobID &  0.501/0.312/0.355 & 0.331/0.182/0.218\\ 
    No\_JobMeta & 0.511/0.338/0.381 & 0.377/0.212/0.253\\
    No\_SeekerMeta &\underline{0.601/0.399/0.449} & 0.505/0.297/0.349\\
    \hline
\end{tabular}
\label{tab:meta}
\end{center}
\end{table}
\setlength{\textfloatsep}{5pt}

\subsection{Negative Sampling Analysis}
\label{sub:negative_sampling_analysis}

We examine the performance of two sampling strategies described in Section~\ref{sub:Experimental Settings} on CB12\_s dataset. According to Table~\ref{tab:strategies}, \textit{PANAP (S1)} outperforms \textit{PANAP (S2)}.
\begin{table}[htbp]
  \caption{Different negative sampling strategies.}
  \begin{center}
  \begin{tabular}{|c|c|}
    \hline
    Sampling\_strategy & HR@5/MRR@5/NDCG@5\\
    \hline
    PANAP (S1) & \textbf{0.756/0.620/0.680}\\
    PANAP (S2) & 0.691/0.492/0.541\\
    \hline
\end{tabular}
\label{tab:strategies}
\end{center}
\end{table}
\setlength{\textfloatsep}{2pt}

\begin{figure}
    \centering
    \begin{subfigure}[b]{0.18\textwidth}
     \centerline{\includegraphics[width=1.\textwidth]{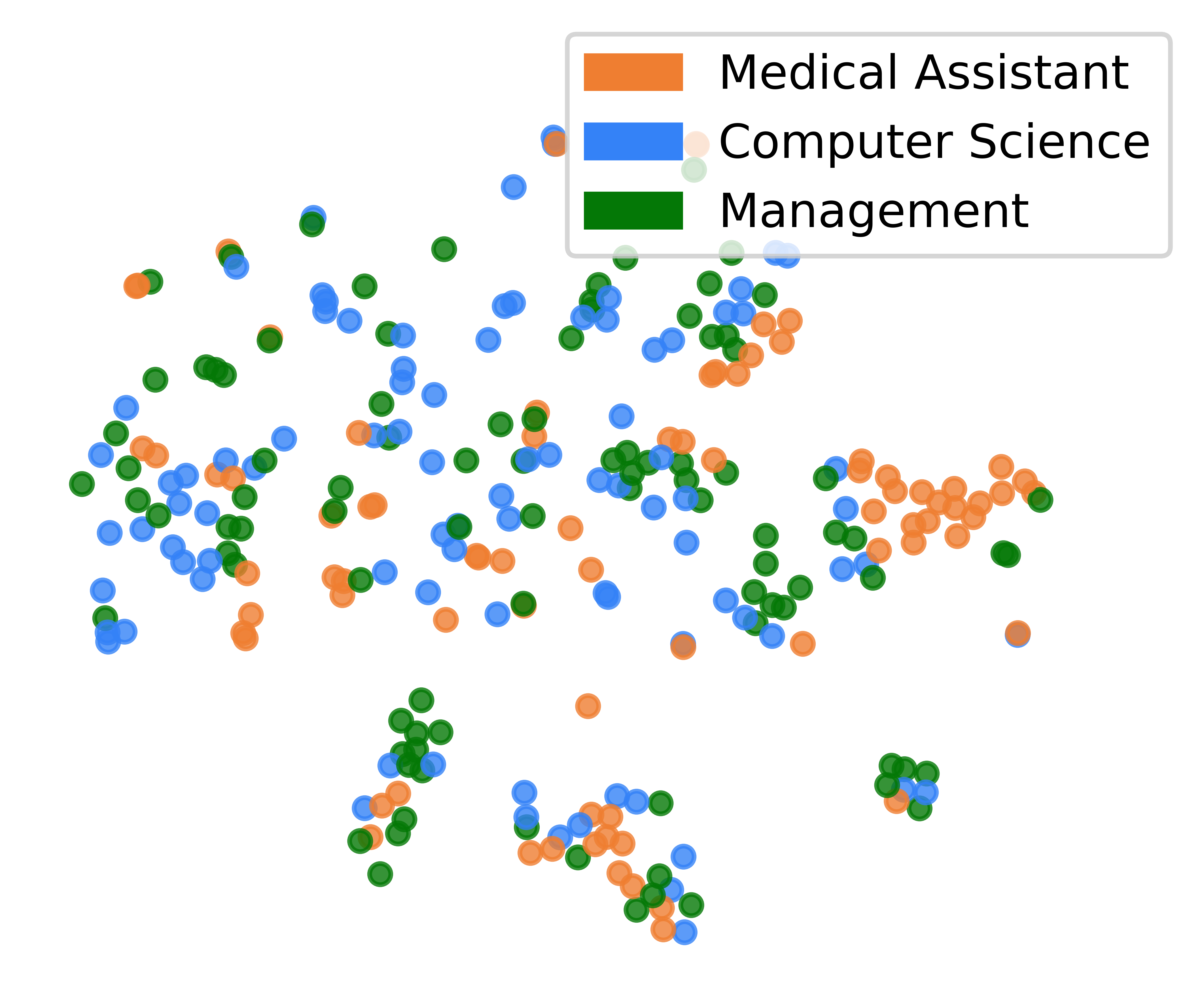}}
    \caption{Strategy 1: \textit{Major}}
    \label{fig:s1_major}
    \end{subfigure}
    \hspace{0.9cm}
    \begin{subfigure}[b]{0.18\textwidth}
     \centerline{\includegraphics[width=1\textwidth]{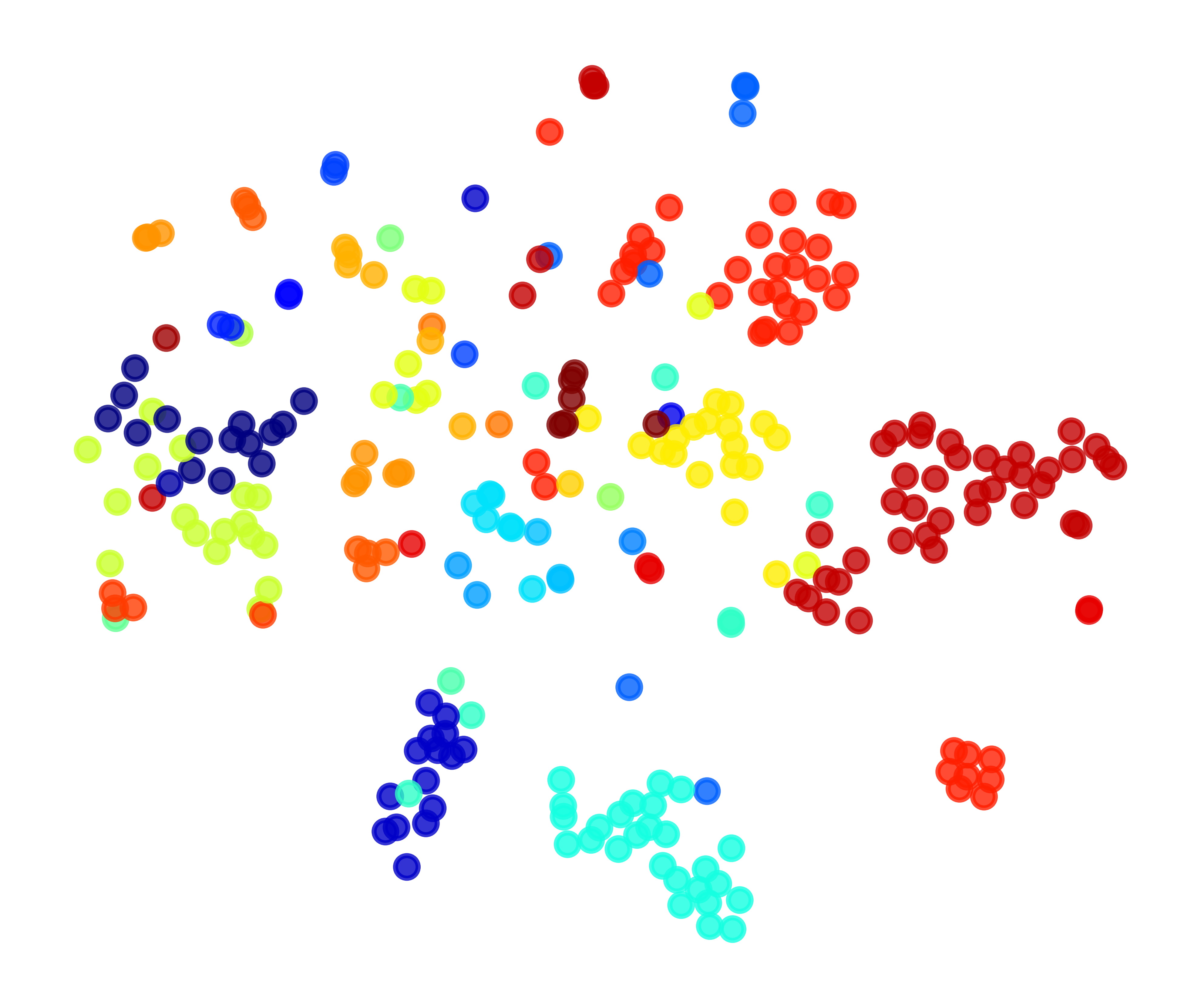}}
    \caption{Strategy 1: \textit{State}}
    \label{fig:s1_state}
  \end{subfigure}
    \begin{subfigure}[b]{0.18\textwidth}
     \centerline{\includegraphics[width=1.\textwidth]{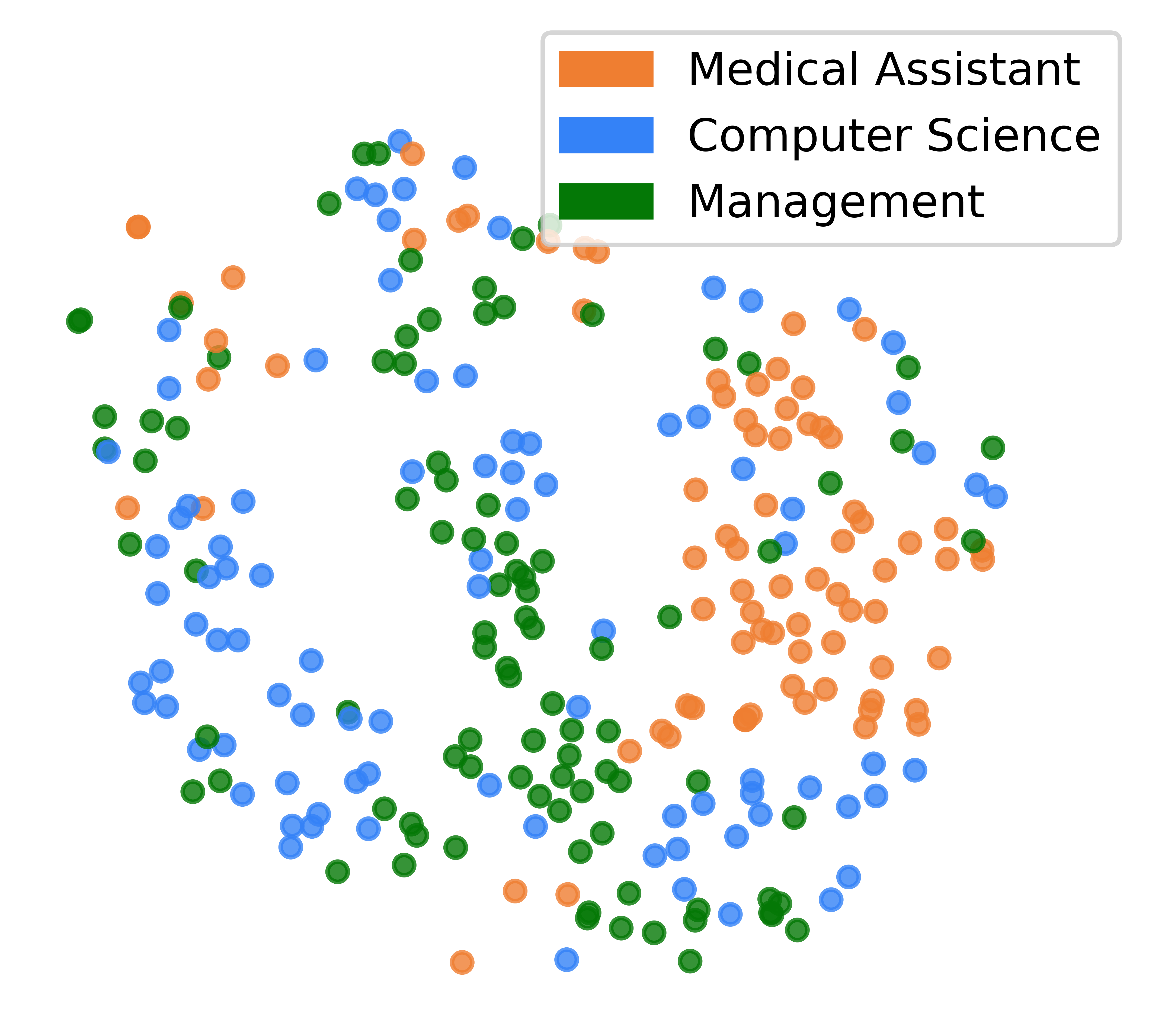}}
    \caption{Strategy 2: \textit{Major}}
    \label{fig:s2_major}
    \end{subfigure}
    \hspace{0.9cm}
    \begin{subfigure}[b]{0.18\textwidth}
     \centerline{\includegraphics[width=1\textwidth]{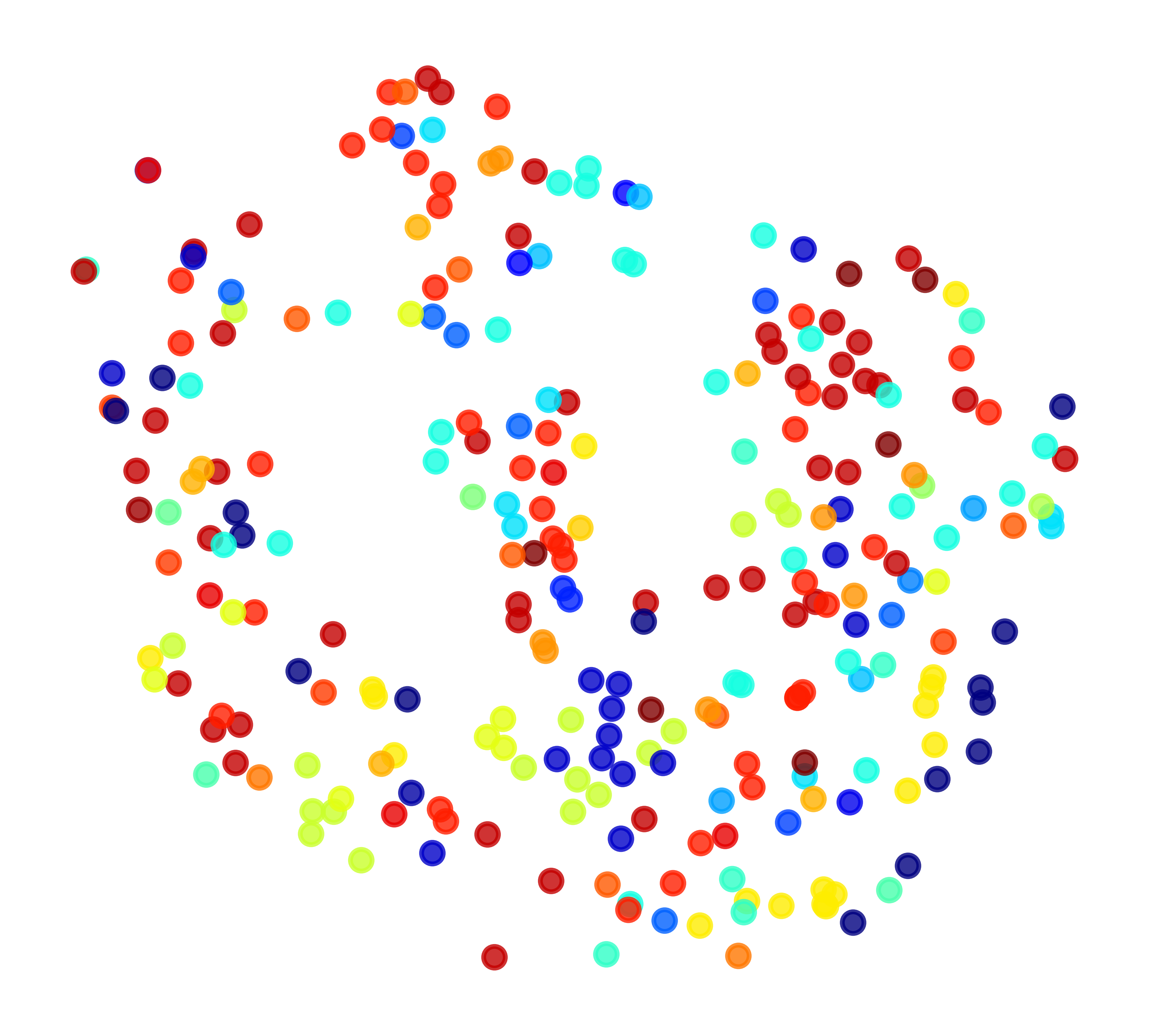}}
    \caption{Strategy 2: \textit{State}}
    \label{fig:s2_state}
  \end{subfigure}
  \caption{Visualization of learned representations.}
  \label{fig:strategy}
\end{figure}

To explain these results, we visualize job seeker representations (session representations) generated from \textit{PANAP (S1)} and \textit{PANAP (S2)} in Figure~\ref{fig:strategy}. We use t-SNE~\cite{van2008visualizing} to reduce representation dimensions. For illustration purposes, we categorize each job seeker according to his/her \textit{Major}. Note that categorizing job seekers through \textit{Major} is not the most reasonable way because some job seekers are currently engaged in occupations that do not match their majors. Thus, we select three non-similar majors. People with these professional backgrounds are more likely to engage in related jobs, including \textit{Management}, \textit{Computer Science} and \textit{Medical Assistant}. We also plot job seeker representations labeled with \textit{State} in Figure~\ref{fig:s1_state} and Figure~\ref{fig:s2_state} to show the influences of different sampling strategies. Each color corresponds to one \textit{State}. We observe from Figure~\ref{fig:s1_state} that representations learned by our model with S1 (i.e., \textit{PANAP (S1)}) are well-clustered into groups, each corresponds to a \textit{State}. This can be used to explain why the accuracy scores of \textit{PANAP (S1)} are better than that of \textit{PANAP (S2)}. A reasonable explanation is that S1 does not consider the ``location'' factor when generating negative samples. More specifically, as mentioned in Section~\ref{sub:Negative Sampling Strategies}, job seekers are more likely to apply for jobs located in their cities or other cities in the same state. The two main reasons why a job seeker does not apply for a job are: the job content is inappropriate, or the work location is not suitable. If most of the negative samples are jobs in other states, regardless of the job content, these negative samples tend to force the model to capture subtle information between different states rather than different job contents. For instance, a job seeker in \textit{Seattle-Washington} has a background of \textit{computer science}. \textit{sales representative} and \textit{java developer} are two negative samples from \textit{Miami-Florida}. Due to the ``location'' factor, this job seeker may not apply for \textit{java developer}, while \textit{sales representative} is not suitable from both ``location'' and job content perspectives. With these negative samples, the model learns to distinguish between locations (e.g., \textit{Seattle-Washington} and \textit{Miami-Florida}) rather than the contents of jobs (e.g., \textit{computer science} and \textit{sales representative}). Therefore, the learned job seeker representations are not categorized according to their \textit{Major} categories, as shown in Figure~\ref{fig:s1_major}. Instead, they are grouped together by the ``location'' as in Figure~\ref{fig:s1_state}. To handle this problem, we propose a location-biased sampling strategy, S2, as described in Section~\ref{sub:Experimental Settings}, which prioritizes jobs in the same state as the job seeker as negative samples. We visualize the representations learned through S2 in Figures~\ref{fig:s2_major} and ~\ref{fig:s2_state} to show the promising results of our proposed strategy. We observe that the job seekers are grouped by \textit{Major} in Figure~\ref{fig:s2_major}, which demonstrates the effectiveness of S2.

\begin{figure}
    \centerline{\includegraphics[width=0.3\textwidth]{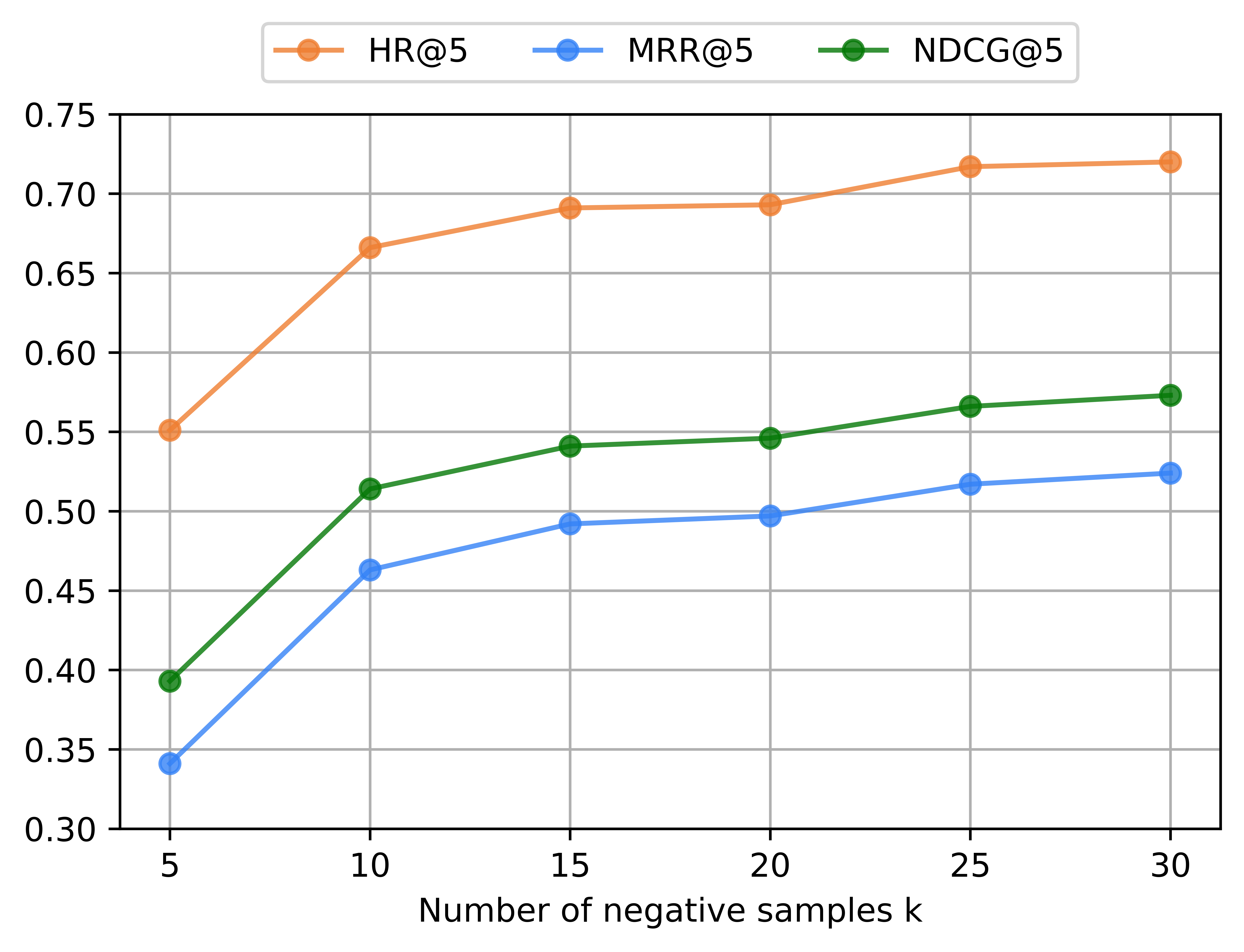}}
    \caption{Different numbers of training negative samples $k$.}
    \label{fig:num_negative}
\end{figure}

\subsection{Number of Negative Samples}
This section discusses the influence of the number of negative samples used for training on prediction accuracy. The experimental results on CB12\_s dataset are shown in Figure~\ref{fig:num_negative}, where $k$ negative samples are used for training and 50 samples for evaluation. When $k$ increases (e.g., from 5 to 10), the performance of our model first has a noticeable improvement. This may be because when $k$ is too small, the information provided by negative samples is relatively limited, and the model cannot learn useful information. Then, when $k$ continues to increase (e.g., from 10 to 20), the performance becomes stable. However, if $k$ is too large, there may not be enough jobs in the same state. Negative samples in other states become dominant, making it difficult for the model to identify the valuable information correctly. As explained in Section~\ref{sub:negative_sampling_analysis}, the model will learn to distinguish the difference of positions, just like \textit{PANAP (S1)}. As a result, the performance consistently improves.

\section{Conclusions}
In this work, we proposed a personalized-attention model for the \textit{Next-Application Prediction} problem, which improves the prediction accuracy. The experiments confirm that, by incorporating personalized-attention, our method can better capture the personal career preference than baseline methods. We investigated the importance of incorporating job content information and metadata in the recruitment domain. Moreover, considering the ``location'' factor when generating negative samples can provide more useful information on job content.

\section*{Acknowledgment}
This work is supported by Randstad corporate research chair in collaboration with Université Paris-Saclay, CentraleSupélec, MICS. We would like to thank the \textit{Mésocentre} computing center of CentraleSupélec and École Normale Supérieure Paris-Saclay~\footnote{\url{http://mesocentre.centralesupelec.fr/}} for providing computing resources.




\vspace{12pt}

\bibliographystyle{IEEEtran}
\bibliography{conference_Jun}
\end{document}